# Numerical Study on Randomization in Late Boundary Layer Transition

Ping Lu[1], Manoj Thapa[2] and Chaoqun Liu[3]

University of Texas at Arlington, Arlington, Texas 76019, USA
cliu@uta.edu

The mechanism of randomization in late boundary layer transition is a key issue of late boundary layer transition and turbulence theory. We studied the mechanism carefully by high order DNS. The randomization was originally considered as a result of large background noise and non-periodic spanwise boundary conditions. It was addressed that the large ring structure is affected by background noises first and then the change of large ring structure affects the small length scale quickly, which directly leads to randomization and formation of turbulence. However, what we observed is that the loss of symmetry starts from the middle level rings while the top and bottom rings are still symmetric. The non-symmetric structure of second level rings will influence the small length scales at the boundary layer bottom quickly. The symmetry loss at the bottom of the boundary layer is quickly spread to up level through ejections. This will lead to randomization of the whole flow field. Therefore, the internal instability of multiple level ring structure, especially the middle ring cycles, is the main reason for flow randomization, but not the background noise. A hypothesis is given that the loss of symmetry may be caused by the shift from C-type transition to K-type transition or reverses.

## Nomenclature

$M_\infty$ = Mach number
$Re$ = Reynolds number
$\delta_{in}$ = inflow displacement thickness
$T_w$ = wall temperature
$T_\infty$ = free stream temperature
$Lz_{in}$ = height at inflow boundary
$Lz_{out}$ = height at outflow boundary
$Lx$ = length of computational domain along x direction
$Ly$ = length of computational domain along y direction
$x_{in}$ = distance between leading edge of flat plate and upstream boundary of computational domain
$A_{2d}$ = amplitude of 2D inlet disturbance
$A_{3d}$ = amplitude of 3D inlet disturbance
$\omega$ = frequency of inlet disturbance
$\alpha_{2d}, \alpha_{3d}$ = two and three dimensional streamwise wave number of inlet disturbance
$\beta$ = spanwise wave number of inlet disturbance
$R$ = ideal gas constant
$\gamma$ = ratio of specific heats
$\mu_\infty$ = viscosity

---

[1] PhD Student, AIAA Student Member, University of Texas at Arlington, USA
[2] PhD Student, AIAA Student Member, University of Texas at Arlington, USA
[3] Professor, AIAA Associate Fellow, University of Texas at Arlington, USA, AIAA Associate Fellow

# I. Introduction

Turbulence is still covered by a mystical veil in nature after over a century of intensive study. Following comments are made by wekipedia web page at http://en.wikipedia.org/wiki/Turbulence : Nobel Laureate Richard Feynman described turbulence as "the most important unsolved problem of classical physics" (USA Today 2006). According to an apocryphal story, Werner Heisenberg was asked what he would ask God, given the opportunity. His reply was: "When I meet God, I am going to ask him two questions: Why relativity? And why turbulence? I really believe he will have an answer for the first." (Marshak, 2005). Horace Lamb was quoted as saying in a speech to the British Association for the Advancement of Science, "I am an old man now, and when I die and go to heaven there are two matters on which I hope for enlightenment. One is quantum electrodynamics, and the other is the turbulent motion of fluids. And about the former I am rather optimistic" (Mullin 1989; Davidson 2004).

These comments clearly show that the mechanism of turbulence formation and sustenance is still a mystery for research. Note that both Heisenberg and Lamb were not optimistic for the turbulence study.

The transition process from laminar to turbulent flow in boundary layers is a basic scientific problem in modern fluid mechanics. After over a hundred of years of study on flow transition, the linear and weakly non-linear stages of flow transition are pretty well understood (Herbert.1988; Kachnaov. 1994). However, for late non-linear transition stages, there are still many questions remaining for research (Kleiser et al, 1991; Sandham et al, 1992; U. Rist et al 1995, Borodulin et al, 2002；Bake et al 2002; Rist et al, 2002; Kachanov, 2003). Adrian (2007) described hairpin vortex organization in wall turbulence, but did not discuss the sweep and ejection events and the role of the shear layer instability. Wu and Moin (2009) reported a new DNS for flow transition on a flat plate. They did obtain fully developed turbulent flow with structure of forest of ring-like vortices by flow transition at zero pressure gradients. However, they did not give the mechanism of the late flow transition. The important mechanism of boundary layer transition such as sweeps, ejections, positive spikes, etc. cannot be found from that paper. Recently, Guo et al (2010) conducted an experimental study for late boundary layer transition in more details. They concluded that the U-shaped vortex is a barrel-shaped head wave, secondary vortex, and is induced by second sweeps and positive spikes. In order to get deep understanding the mechanism of the late flow transition in a boundary layer and physics of turbulence, we recently conducted a high order direct numerical simulation (DNS) with 1920×128x241 gird points and about 600,000 time steps to study the mechanism of the late stages of flow transition in a boundary layer at a free stream Mach number 0.5 (Chen et al., 2009, 2010a, 2010b, 2011a, 2011b; Liu et al., 2010a, 2010b, 2010c, 2011a, 2011b 2011c, 2011d; Lu et al., 2011a, 2011b). The work was supported by AFOSR, UTA, TACC and NSF Teragrid. A number of new observations are made and new mechanisms are revealed in late boundary layer transition.

Randomization is a key issue of late boundary layer transition and turbulence formation. This work is devoted to the investigation of the late stages of the laminar-turbulent transition process in a flat-plate boundary layer. As we know, in order to get a fully developed turbulent flow, the following two characteristics should be obtained: 1) small size vortices; 2) randomization. There are not many existing literatures investigating the mechanism of randomization. Here, we only take those conclusions into account, which were made by Meyer and his co-workers (see Meyer et al 2003). They believe that "the inclined high-shear layer between the legs of the $\Lambda$-vortex exhibits increasing phase jitter (i.e randomization) starting from its tip towards the wall region." However, by carefully checkin our DNS data, we observed a phenomenon which is different from the hypothesis given by Meyer and his co-workers.

A $\lambda 2$ technology developed by Jeong and Hussain (1995) is used for visualization.

## II. Case Setup and Code Validation

### 2.1 Case setup
The computational domain is displayed in Figure 1. The grid level is 1920×128×241, representing the number of grids in streamwise ($x$), spanwise ($y$), and wall normal ($z$) directions.

The grid is stretched in the normal direction and uniform in the streamwise and spanwise directions. The length of the first grid interval in the normal direction at the entrance is found to be 0.43 in wall units ($Y^+$=0.43). The parallel computation is accomplished through the Message Passing Interface (MPI) together with domain decomposition in the streamwise direction (Figure 2). The flow parameters, including Mach number, Reynolds number, etc are listed in Table 1. Here, $x_{in}$ represents the distance between leading edge and inlet, $Lx$, $Ly$, $Lz_{in}$ are the lengths of the computational domain in x-, y-, and z-directions, respectively, and $T_w$ is the wall temperature.

Table 1: Flow parameters

| $M_\infty$ | $Re$ | $x_{in}$ | $Lx$ | $Ly$ | $Lz_{in}$ | $T_w$ | $T_\infty$ |
|---|---|---|---|---|---|---|---|
| 0.5 | 1000 | 300.79$_{in}$ | 798.03$_{in}$ | 22$_{in}$ | 40$_{in}$ | 273.15K | 273.15K |

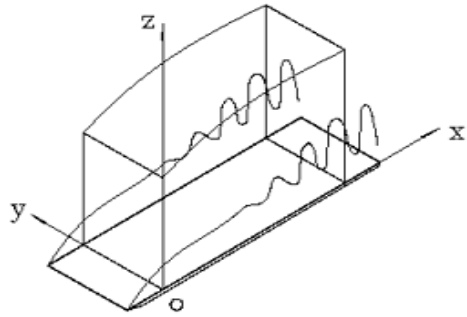

Figure 1: Computation domain

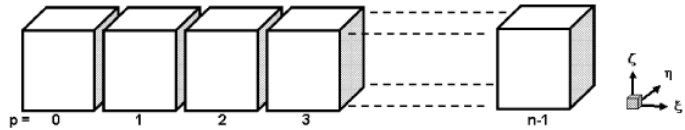

Figure 2: Domain decomposition along the streamwise direction in the computational space

**2.2 Code Validation**

The DNS code – "DNSUTA" has been validated by NASA Langley and UTA researchers (Jiang et al, 2003; Liu et al, 2010a; Lu et al 2011b) carefully to make sure the DNS results are correct.

**2.2.1 Comparison with Linear Theory**

Figure 3 compares the velocity profile of the T-S wave given by our DNS results to linear theory. Figure 4 is a comparison of the perturbation amplification rate between DNS and LST. The agreement between linear theory and our numerical results is quite good.

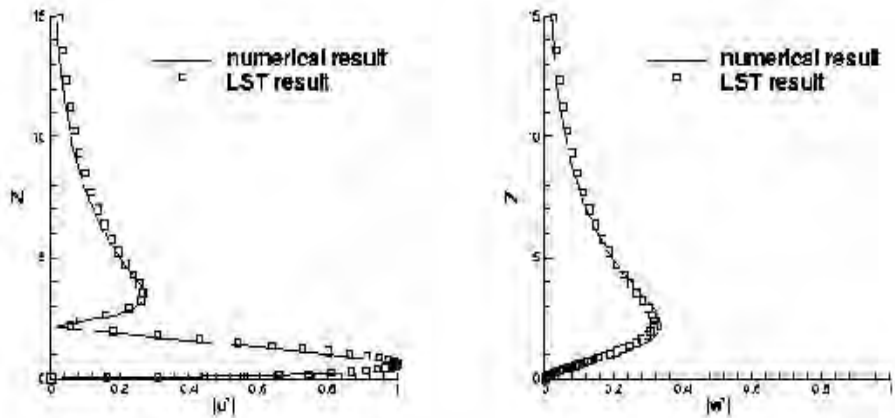

Figure 3: Comparison of the numerical and LST velocity profiles at Rex=394300

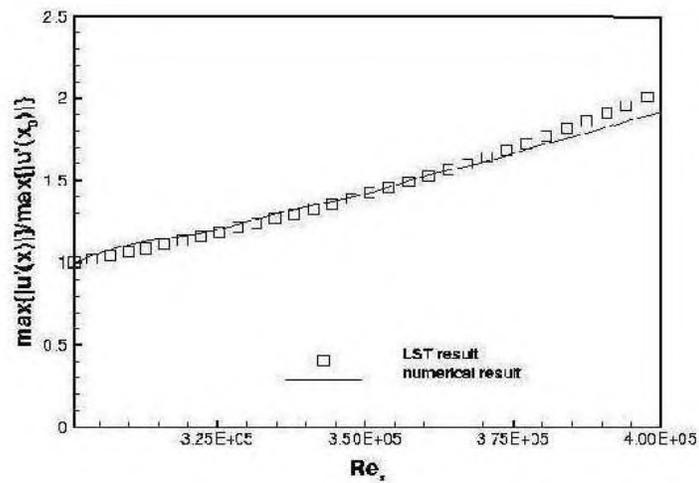

Figure 4: Comparison of the perturbation amplification rate between DNS and LST

**2.2.2 Grid Convergence**

The skin friction coefficient calculated from the time-averaged and spanwise-averaged profile on a coarse and fine grid is displayed in Figure 5. The spatial evolution of skin friction coefficients of laminar flow is also plotted out for comparison. It is observed from these figures that the sharp growth of the skin-friction coefficient occurs after $x \approx 450_{in}$, which is defined as the 'onset point'. The skin friction coefficient after transition is in good agreement with the flat-plate theory of turbulent boundary layer by Cousteix in 1989 (Ducros, 1996). Figures 5(a) and 5(b) also show that we get grid convergence in skin friction coefficients.

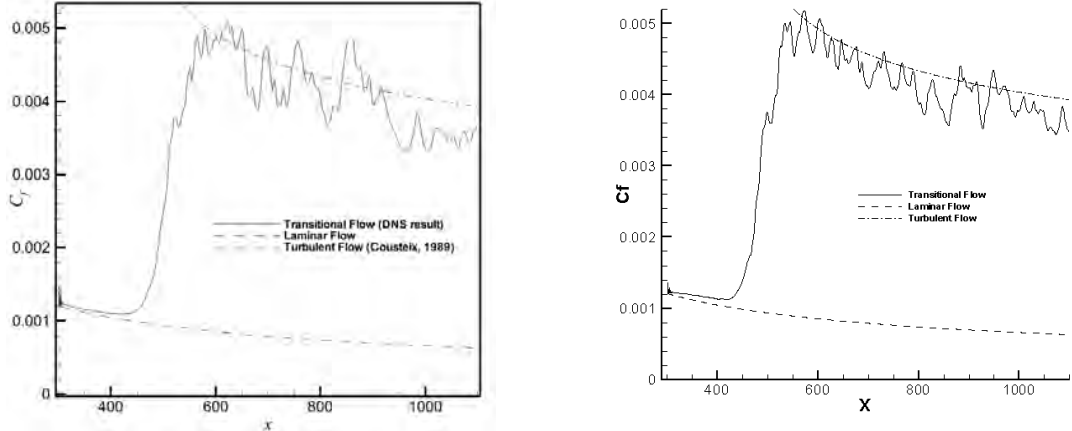

(a) Coarse Grids ($960\times 64\times 121$)  (b) Fine Grids (1920x128x241)
Figure 5: Streamwise evolutions of the time-and spanwise-averaged skin-friction coefficient

### 2.2.3 Comparison with Log Law

Time-averaged and spanwise-averaged streamwise velocity profiles for various streamwise locations in two different grid levels are shown in Figure 6. The inflow velocity profiles at $x = 300.79_{in}$ is a typical laminar flow velocity profile. At $x = 632.33_{in}$, the mean velocity profile approaches a turbulent flow velocity profile (Log law). This comparison shows that the velocity profile from the DNS results is turbulent flow velocity profile and the grid convergence has been realized.

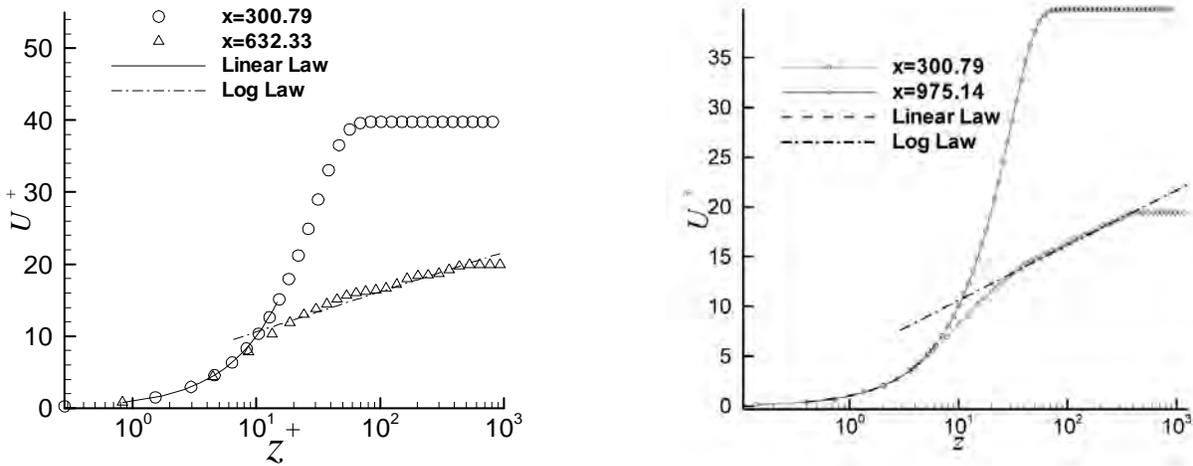

(a) Coarse Grids (960x64x121)  (b) Fine Grids (1920x128x241)
Figure 6: Log-linear plots of the time-and spanwise-averaged velocity profile in wall unit

### 2.2.4 Spectra and Reynolds stress (velocity) statistics

Figure 7 shows the spectra in x- and y- directions. The spectra are normalized by z at location of $Re_x = 1.07\times 10^6$ and $y^+ = 100.250$. In general, the turbulent region is approximately defined by $y^+ > 100$ and $y/\delta < 0.15$. In our case, The location of $y/\delta = 0.15$ for $Re_x = 1.07\times 10^6$ is corresponding to $y^+ \approx 350$, so the points at $y^+ = 100$ and $250$ should be in the turbulent region. A straight line with slope of -3/5 is also shown for comparison. The spectra tend to tangent to the $k^{-\frac{5}{3}}$ law.

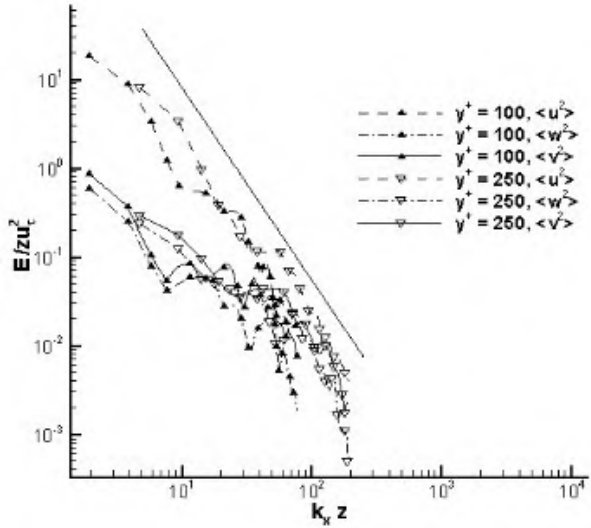 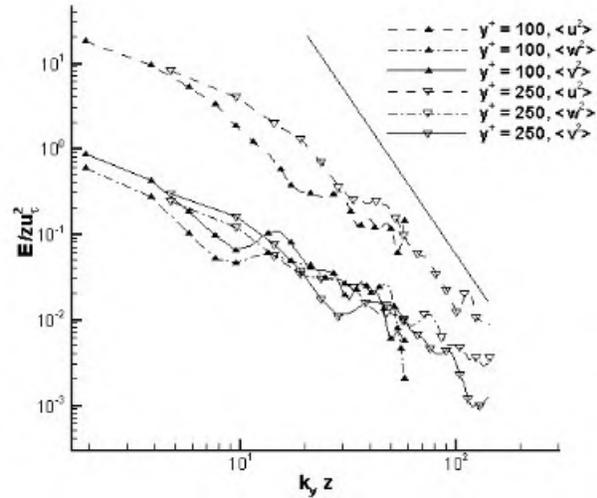

Figure 7(a): Spectra in x- direction  Figure 7(b): Spectra in y- direction

Figure 8 shows Reynolds shear stress profiles at various streamwise locations, normalized by square of wall shear velocity. There are 10 streamwise locations starting from leading edge to trailing edge are selected. As expected, close to the inlet at $Re_x = 326.8 \times 10^3$ where the flow is laminar, the values of the Reynolds stress is much smaller than those in the turbulent region. The peak value increases with the increase of $x$. At around $Re_x = 432.9 \times 10^3$, a big jump is observed, which indicates the flow is in transition. After $Re_x = 485.9 \times 10^3$, the Reynolds stress profile becomes close to each other in the turbulent region. So for this case, we can consider that the flow transition starts after $Re_x = 490 \times 10^3$.

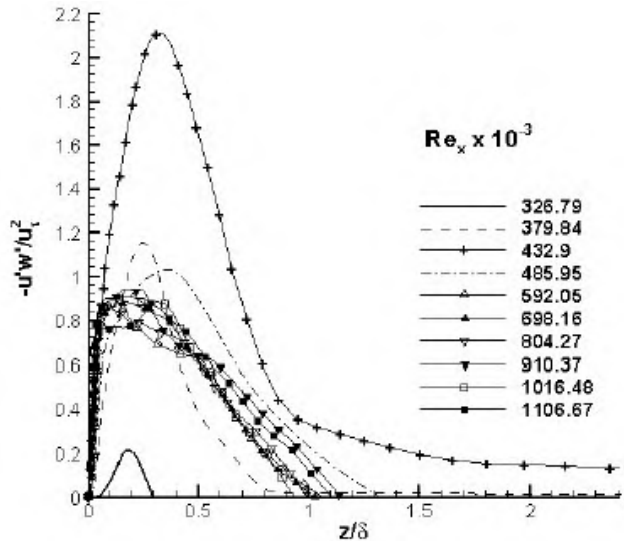

Figure 8: Reynolds stress

### 2.2.5 Comparison with Other DNS

Although we cannot compare our DNS results with those given by Borodulin et al (2002) quantitatively, we still can find that the shear layer structure are very similar in two DNS computations in Figure 9.

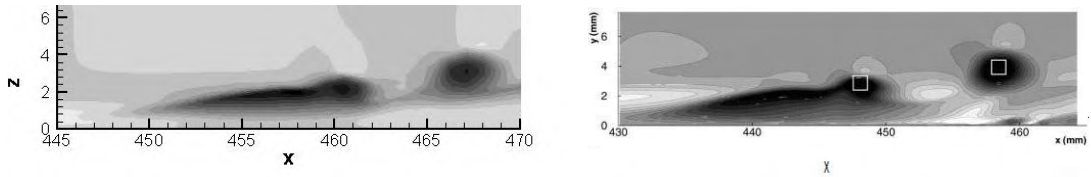

(a) Our DNS  (b) Borodulin et al (2002)

Figure 9: Qualitatively Comparison of contours of streamwise velocity disturbance u in the (x, z)-plane (Light shades of gray correspond to high values)

## 2.2.6 Comparison with Experiment

By this $\Lambda_2$-eigenvalue visualization method, the vortex structures shaped by the nonlinear evolution of T-S waves in the transition process are shown in Figure 10. The evolution details are briefly studied in our previous paper (Chen et al 2009) and the formation of ring-like vortices chains is consistent with the experimental work (Lee C B & Li R Q, 2007, Figure 11) and previous numerical simulation by Rist and his co-authors (Bake et al 2002).

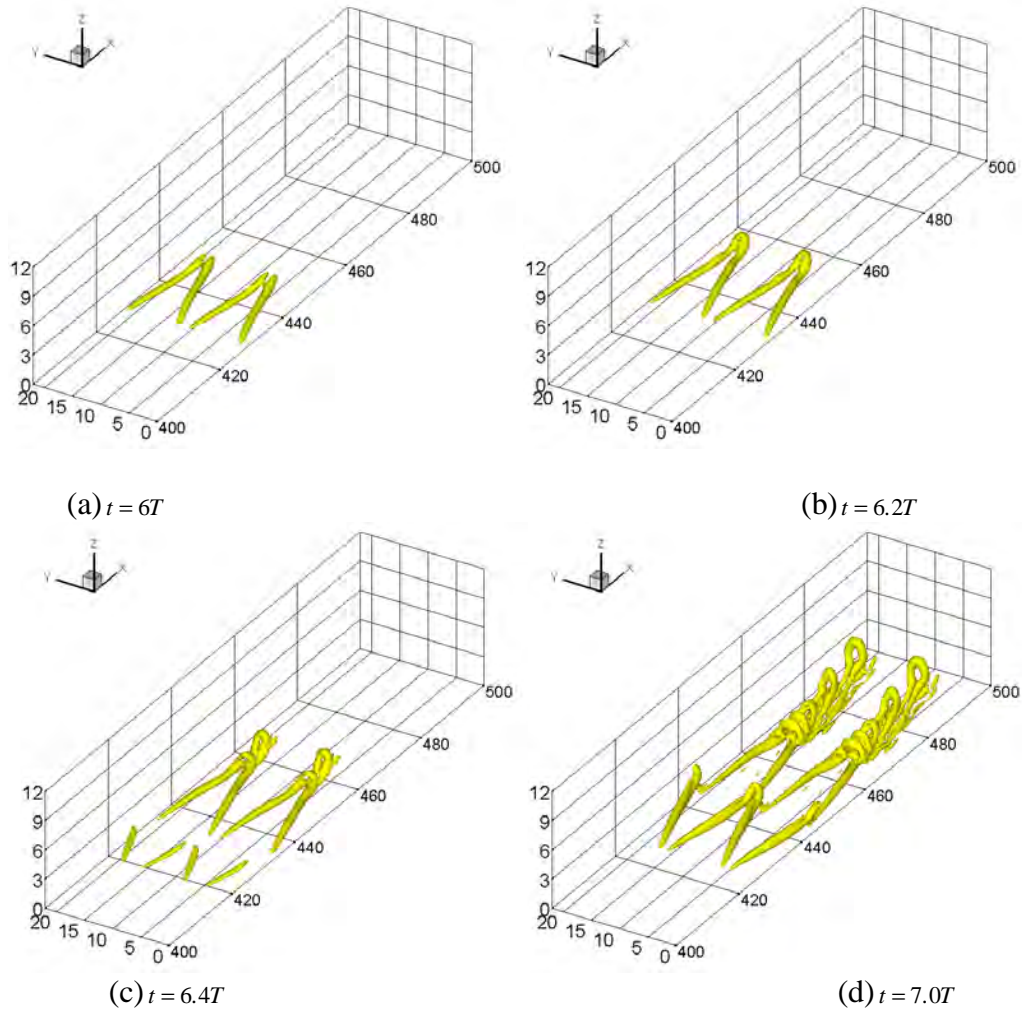

(a) $t = 6T$  (b) $t = 6.2T$

(c) $t = 6.4T$  (d) $t = 7.0T$

Figure 10: Evolution of vortex structure at the late-stage of transition
(Where $T$ is the period of T-S wave)

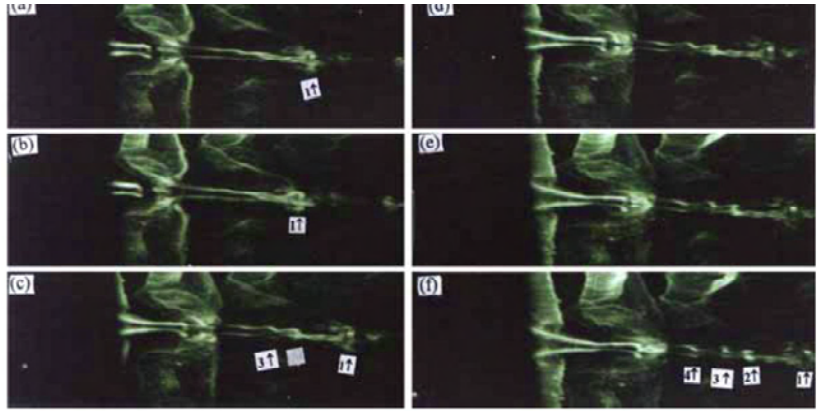

Figure 11: Evolution of the ring-like vortex chain by experiment (Lee et al, 2007)

**2.2.7 U-shaped vortex in comparison with experimental results**

Figure 12(a) (Guo et al, 2010) represents an experimental investigation of the vortex structure including ring-like vortex and barrel-shaped head (U-shaped vortex). The vortex structures of the nonlinear evolution of T-S waves in the transition process are given by DNS in Figure 12(b). By careful comparison between the experimental work and DNS, we note that the experiment and DNS agree with each other in a detailed flow structure comparison. This cannot be obtained by accident, but provides the following clues: 1) Both DNS and experiment are correct 2) Disregarding the differences in inflow boundary conditions (random noises VS enforced T-S waves) and spanwise boundary conditions (non-periodic VS periodic) between experiment and DNS, the vortex structures are same 3) No matter K-, H- or other types of transition, the final vortex structures are same 4) There is an universal structure for late boundary layer transition 5) turbulence has certain coherent structures (CS) for generation and sustenance.

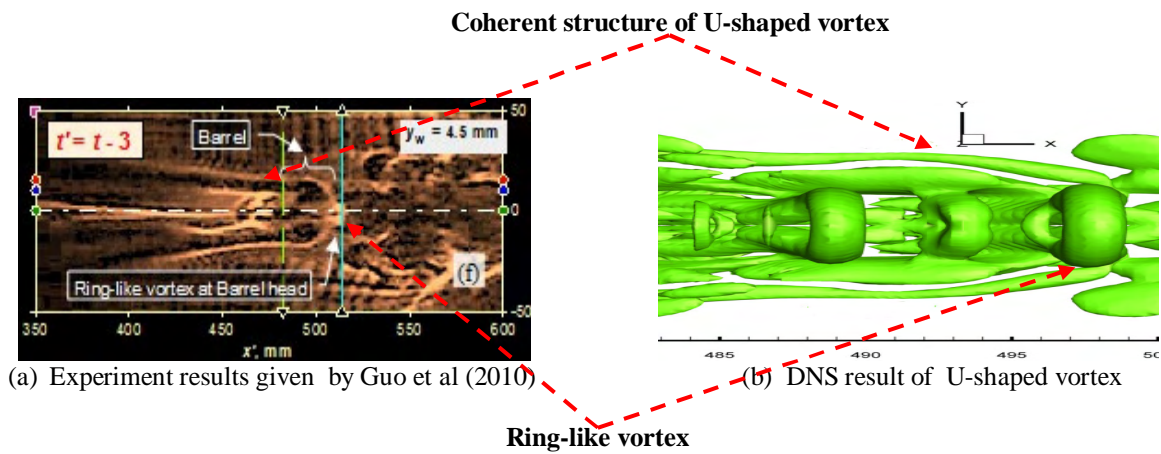

(a) Experiment results given by Guo et al (2010)  (b) DNS result of U-shaped vortex

Figure 12: Qualitative vortex structure comparison with experiment

**All these verifications and validations above show that our code is correct and our DNS results are reliable.**

# III Our DNS Observations on "Randomization"

## 3.1 Asymmetric phenomenon happened at the middle level in both streamwise and spanwise directions

As we described (Liu et al 2010a, 2010b, 2011a; Lu et al, 2011b), the instability of shear layer is the origin of the vortex ring generation at the late stages of boundary layer transition. We first look at the vortex structure at the time step of 15.5 T from the top (Figure 13a) and bottom (Figure 13b) which were visualized by using $\Lambda_2$ method. From the top view, we observe that the large vortex structure still keeps symmetric (Figure 13a). Meanwhile, one cross section in Figure 13b is selected to investigate the mechanism of randomization along the spanwise direction. By checking the streamtrace of each vortex around the boundary layer (Figures 14 and 15), the following phenomena have been observed:

1. In the red frames in Figures 14 and 15, the left vortex ring is generated around the shear layer, in which the velocity is changed from the green level to yellow level. On the other hand, the right vortex ring is generated by a shear layer between the red and yellow levels. However, all of the other vortex rings are remaining almost symmetric on both sides.
2. It is also found that the loss of symmetry happens in the middle of the flow field in the streamwise direction, not inflow and not outflow. Since all noises are mainly introduced through the inflow, outflow or far field, it is unlikely that the reason to cause asymmetry is due to the large background noises, but is pretty much the internal property of the multiple level vortex structure in boundary layers
3. The above observation demonstrates that vortex rings at the middle level in the normal direction lose the symmetry first but not on the top or bottom level rings. Figure 13(a) shows that the structure of the top vortex rings still keep symmetric as before although the middle level rings have lost the symmetry.
4. When the rings at the middle part loss the symmetry, it will quickly affect the shape of rings at the bottom of boundary layer through sweep motion (downdraft) as shown in Figures 14 and 15.

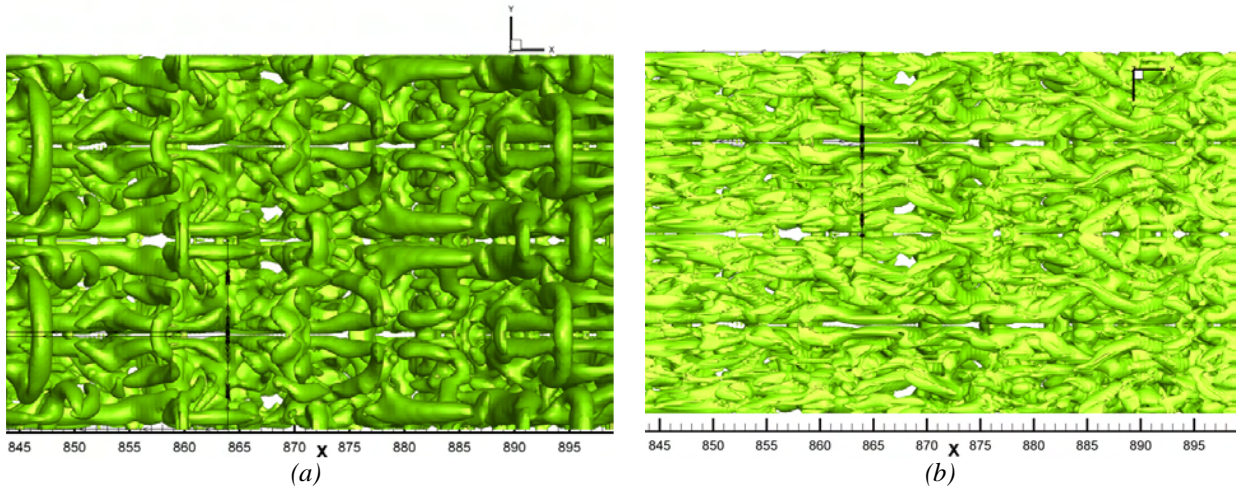

*(a)*        *(b)*

Figure 13: top and bottom view-isosurface of $\Lambda_2$

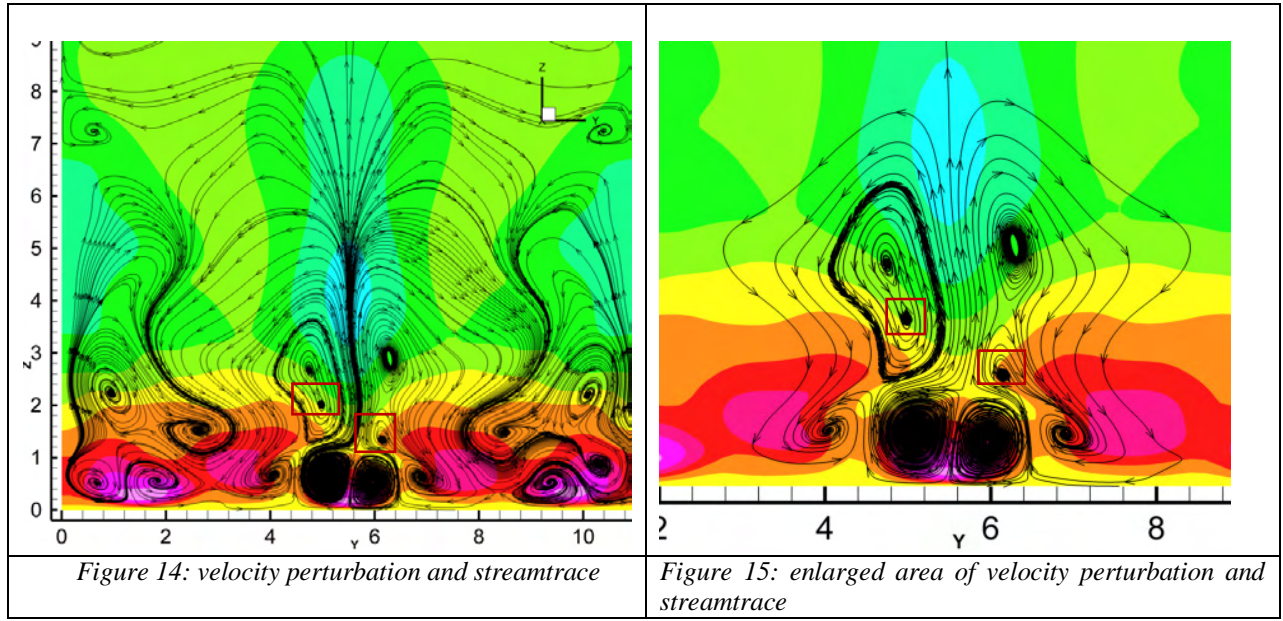

Figure 14: velocity perturbation and streamtrace

Figure 15: enlarged area of velocity perturbation and streamtrace

**3.2 The evolution of shear layers**

The sweeps generated by the vortex rings bring high speed momentum from the inviscid area to the lower boundary layer (Liu et al, 2010b, 2011d; Lu et al, 2011b). Due to the large difference of streamwise velocity, the high shear layer regions can be recognized near the wall region. In order to show the 3-D characteristics of the high shear layer, it is helpful to check the parameter variation at different time steps (t=16.875T, t=17.75T, t=24.75T). The spanwise vorticity $\omega_y$ is checked. We use the positive values of the wall-normal velocity gradient $\partial U / \partial z$ plotted as contour lines in Figure 16, where the yellow area represents the high value of vorticity. In boundary layers, the gradient $\partial U / \partial z$ is almost equal to the spanwise vorticity $\omega_y$ because $\partial V / \partial x$ is rather small (see e.g. Rist & Fasel 1995). From these figures, following results can be obtained:

1) As we addressed, the vortex rings were generated by different strength of streamwise velocity shear layers. From those figures at different time steps, these shear layers are not symmetric anymore. This asymmetry consequently affects the shape of the ring-like vortex due to the change of the direction of sweeps (downdraft motion). As a result, the intensity of positive spike is reduced by one side. That is reason why we can see asymmetric high shear layer in the near wall region inside the red frame.
2) At time step t=17.75T in Figure16 (b), it can be clearly seen that the middle part of shear layer inside the red frame is getting weaker than one at t=16.875T in Figure 16(a).
3) We have selected a same value for iso-surface of $\omega_y$ at the three different time steps. Finally, Figure16(c) indicates that there are no high shear layers occurred at the middle part of the flow field. This means the intensity of shear layer is very low at t=24.75T.

We have addressed that the intensity of shear layer at the bottom is getting weaker and weaker as flow travels downstream. This phenomenon can be explained as follows: While the ring-like vortices travel downstream, they will be lifted up due to the boundary layer mean velocity profile. Since the shear layer is caused by sweeps generated by ring-like vortices which are lifted up. Then, the height of the shear layer becomes much higher than the past, changing from 0.3 to 0.5 and 1.0 at three different time steps (Figure 17 (a)-(c)). On the other hand, while more and more small vortices are generated near the wall region, we can find that the value of velocity perturbation is enhanced when the flow travels downstream.

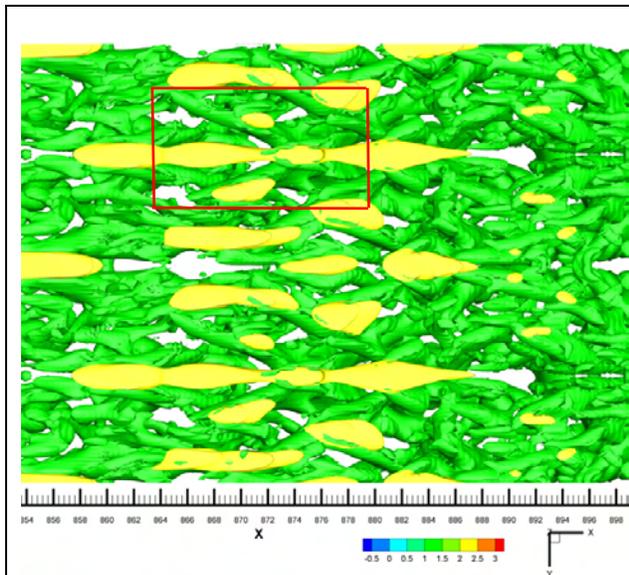

*Figure 16(a): isosurface of $\Lambda_2$ and shear layer at t=16.875T*

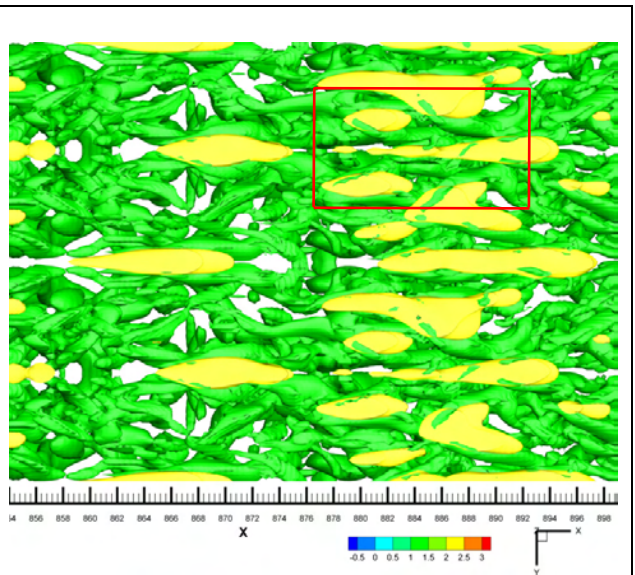

*Figure 16(b): isosurface of $\Lambda_2$ and shear layer at t=17.75T*

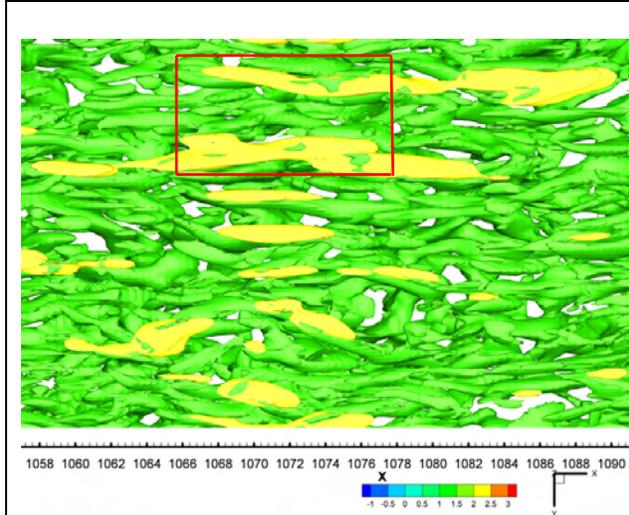

*Figure 16(c) isosurface of $\Lambda_2$ and shear layer at t=24.75T*

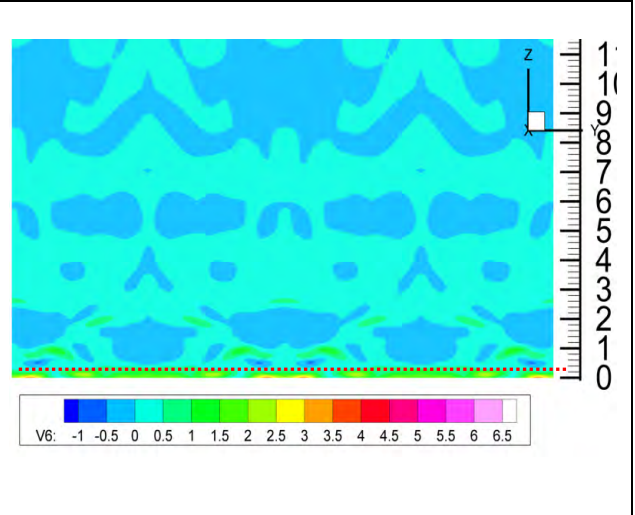

*Figure 17(a). cross-section with shear layer at t=16.5T*

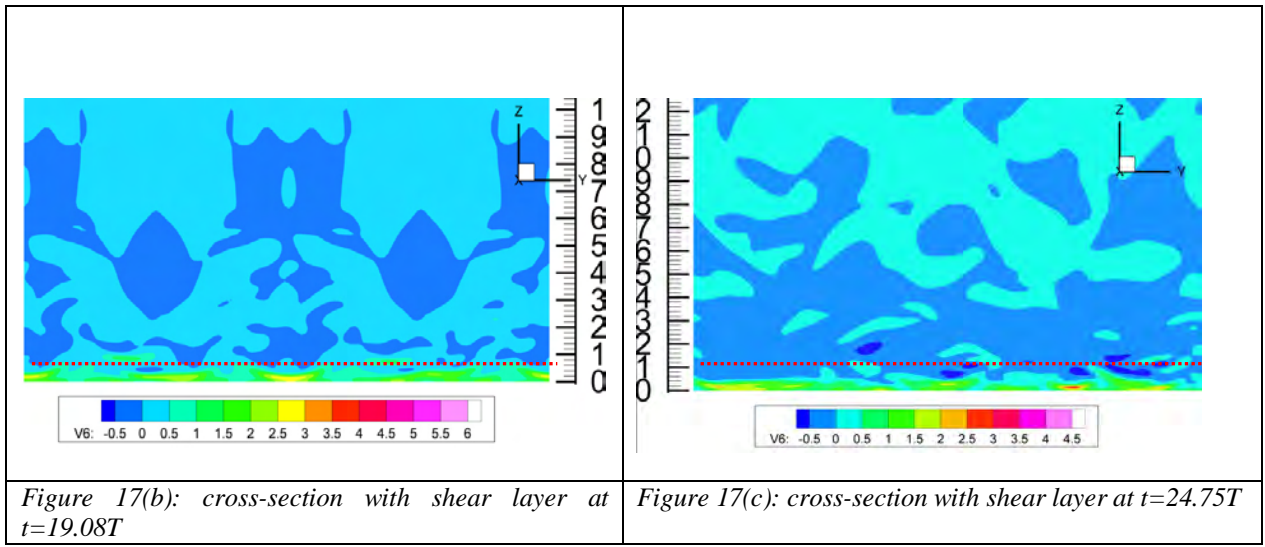

| Figure 17(b): cross-section with shear layer at t=19.08T | Figure 17(c): cross-section with shear layer at t=24.75T |

## 3.3 Completely asymmetric flow at very late stages

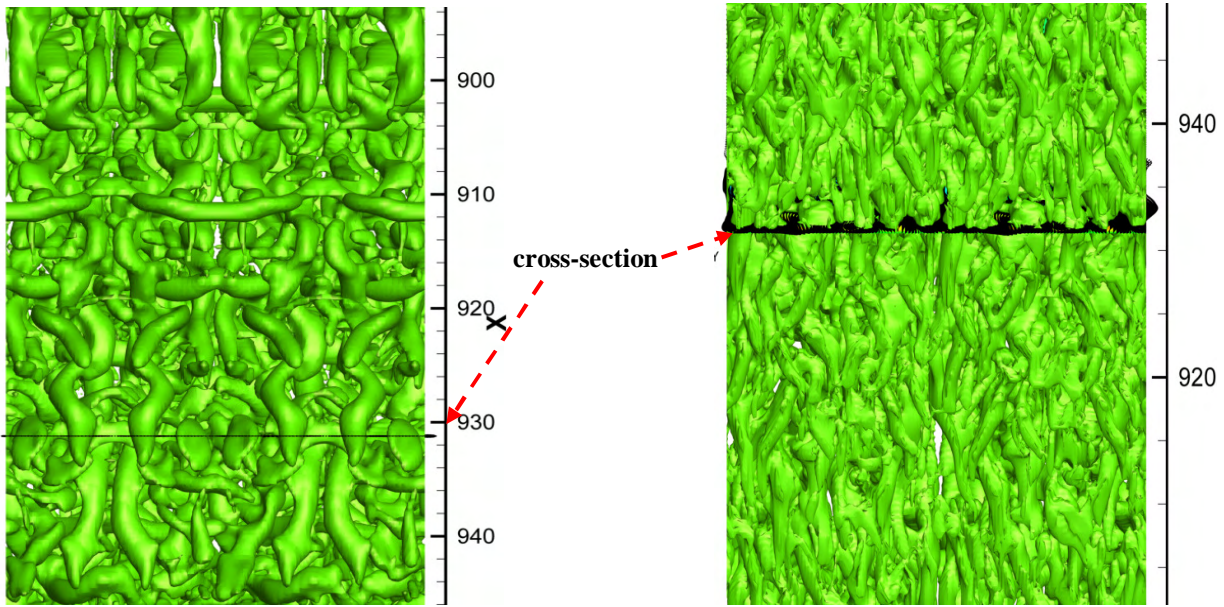

Figure 18(a): Top view - isosurface of $\Lambda_2$     Figure 18(b): Bottom view - isosurface of $\Lambda_2$

Another evidence to prove that the asymmetric phenomenon happened in the middle level of vortex ring cycles can be seen at the time step t= 18.625T. By looking at the top view of the flow structure (Figure18 (a)), we note that the top vortex structure is still keeping symmetric. However, the bottom structure has totally lost symmetric characteristics (Figure 18(b)). The conclusion is that the asymmetric fluctuations did not yet spread very far to the top in the normal direction away from the place where the original asymmetry. This indicates that the mechanism of growth of asymmetry is related to the development of those deformed vortices in the boundary layer by ejection (upward) motion. As the time develops, there are other asymmetric displacements of the vortices. Then asymmetric phenomena can be observed for almost every vortex in Figure 18(c) which is one cross-section taken from Figures 18(a) and 18(b) at time step t=18.625T. Finally, in addition to the boundary layer, the whole flow field is filled with many vortices with asymmetry at time t= 25.0T. This can be seen in Figure 19(a) which is one cross-section from Figure 19(b). This can be seen even better in the top visualization in Figure 19(b), which indicates the flow has become fully developed turbulent flow.

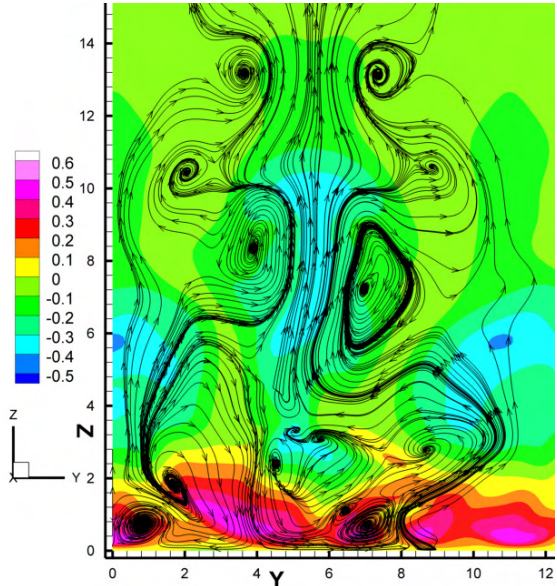

*Figure 18(c): velocity perturbation with streamtrace*

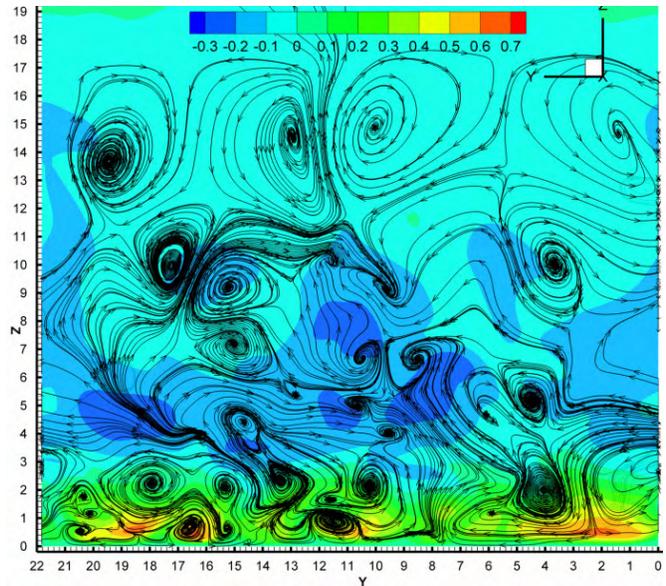

*Figure 19(a): velocity perturbation with streamtrace*

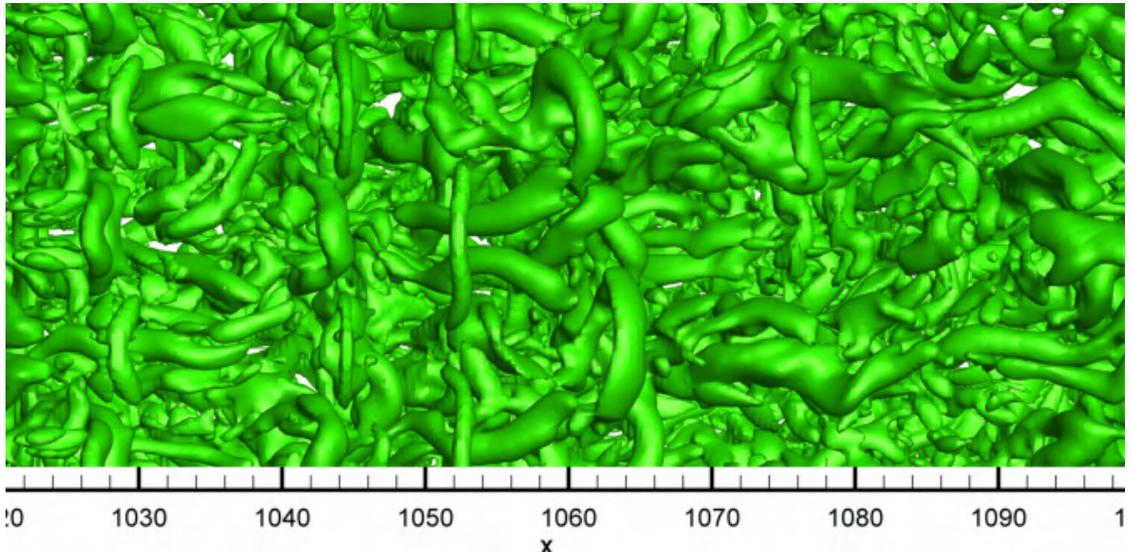

*Figure 19(b): Top view-isosurface of $\Lambda_2$*

## IV. Analysis

The flow is symmetric and periodic with period=π (Figure 20) at early stages. Since we enforce periodic boundary condition in the spanwise direction with a period=2π and enforce a symmetric and periodic (period=π) flow profile at the inflow, the flow will have to keep periodic in the spanwise direction (period=2π) and symmetric at inflow. However, the following processes have been observed: 1) Flow lost symmetry in the middle (not at inflow or outflow) in the streamwise direction and the middle of the multiple overlapping ring cycles (Figure 21). However, the flow is still periodic with a period=π (Figures 22(a) and 22(b)). This means the flow does not only have $\sum_{k=0}^{n} a_k \cos(2ky)$ but also have $\sum_{k=0}^{n} b_k \sin(2ky)$ which is newly generated; 2) Flow lost periodicity with period=π, but has to be periodic with period=2π (Figure 22(c) and 22(d)), which we enforced. Since the DNS study is focused on the mechanism of randomization and the DNS computation only allows use two periods in the spanwise direction, we consider the flow is randomized when the symmetry is lost and period is changed from π to 2π (Figure 23(a) and (b)):

$$f(y) = \sum_{k=0}^{n} a_k \cos(2ky) + \sum_{k=0}^{n-1} b_k \sin(2ky) + \sum_{k=0}^{n-1} c_k \cos(2ky + y) + \sum_{k=0}^{n-1} d_k \sin(2ky + y)$$

In real flow, there is no such a restriction of periodic boundary condition in the spanwise direction.

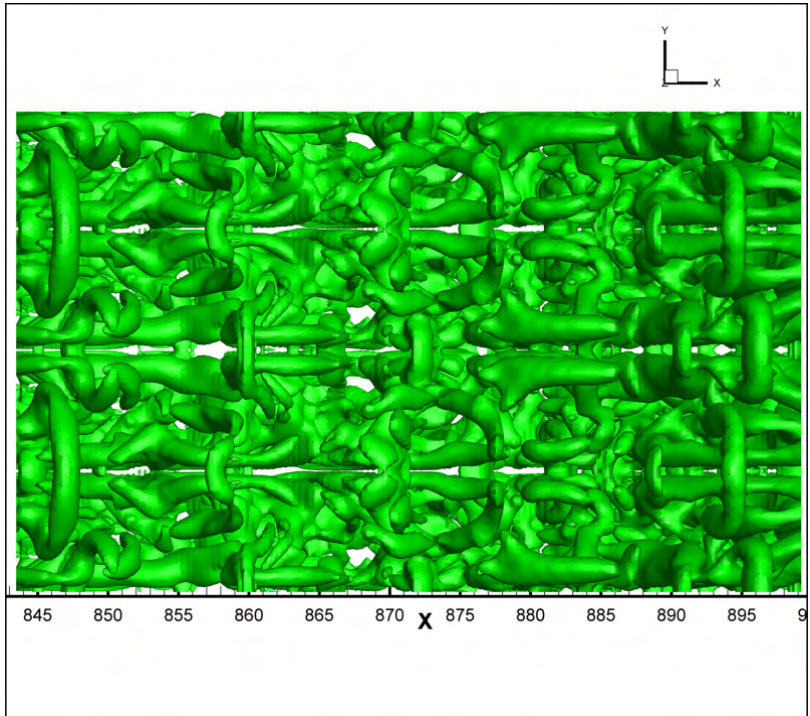

Figure 20: Whole domain is symmetric and periodic: f(-y)=f(y) and f(y+π)=f(y)
(the period=π; spanwise domain is –π<y<π)

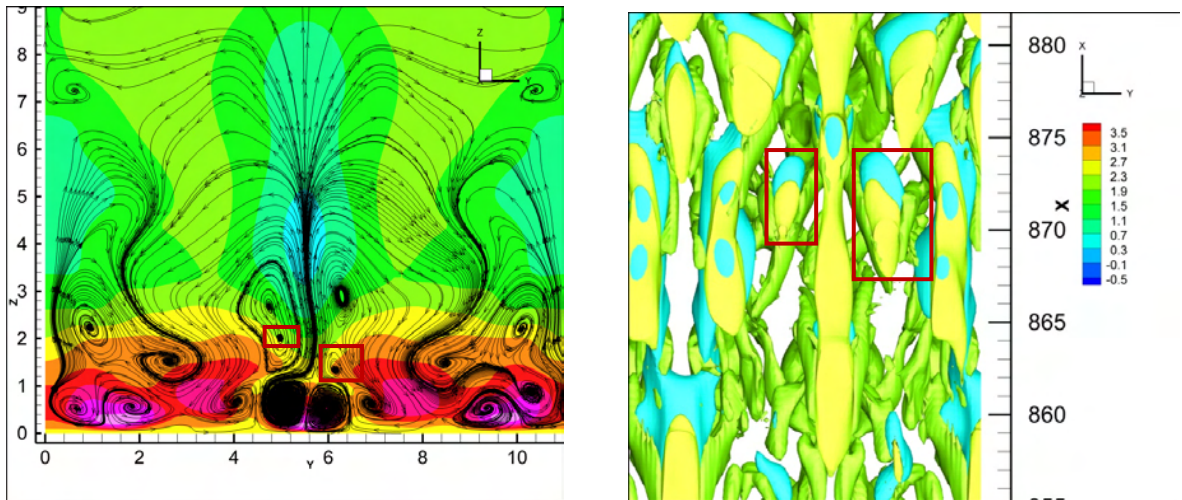

(a) Section view in y-z plane  (b) Bottom view of positive spike

Figure 21: The flow lost symmetry in second level rings and bottom structure at T=15.0

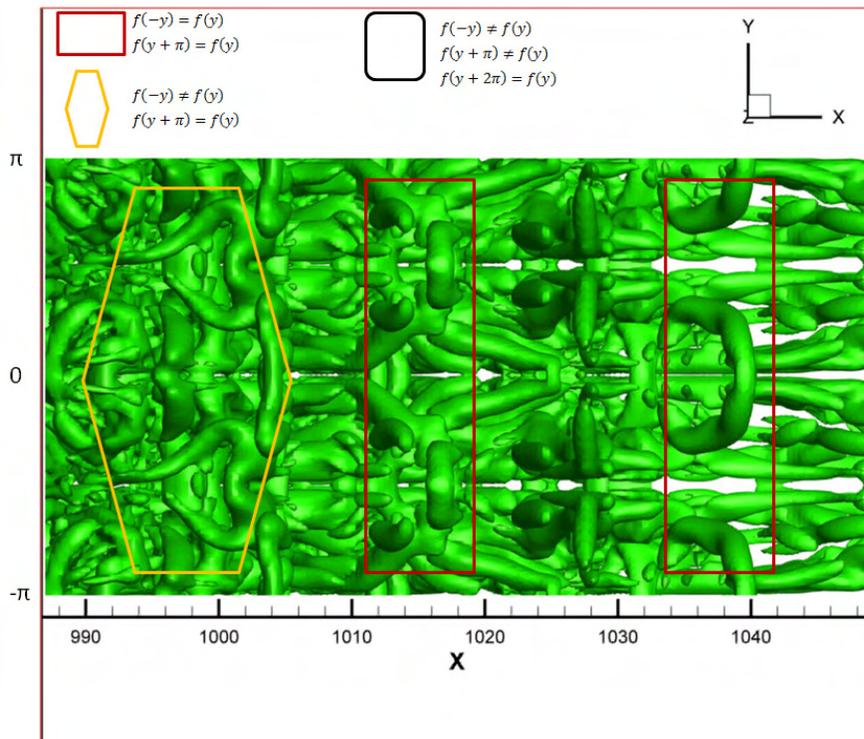

Figure 22(a): Symmetric and asymmetric at t=24.3T – Red rectangular frame: periodic and symmetric at y=-π/2, 0, π/2, i.e. f(y+π)=f(y), f(-π/2 –y)=f(-π/2+y), f(-y)=f(y), f(π/2-y)=f(π/2+y); Yellow diamond frame: periodic, f(y+π)=f(y), period=π; but asymmetric $f(-y) \neq f(y)$; the spanwise domain is –π<y<π

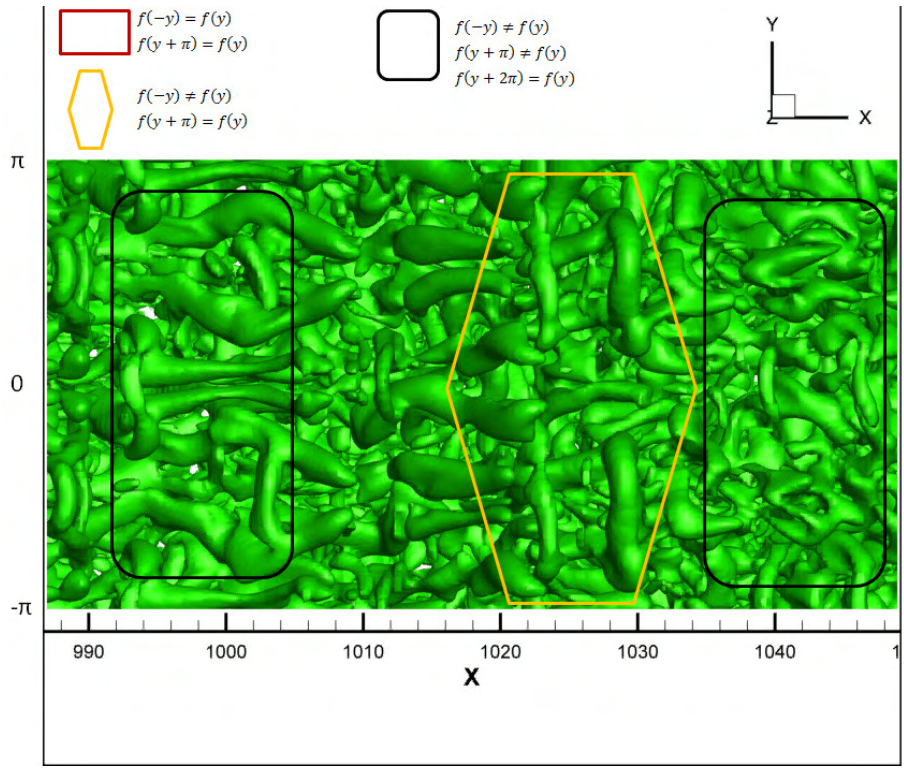

Figure 22(b): Periodic but asymmetric at t=26.5T – Yellow diamond frame: periodic, period=π; black box: periodic but period= 2π; the spanwise domain is –π<y<π

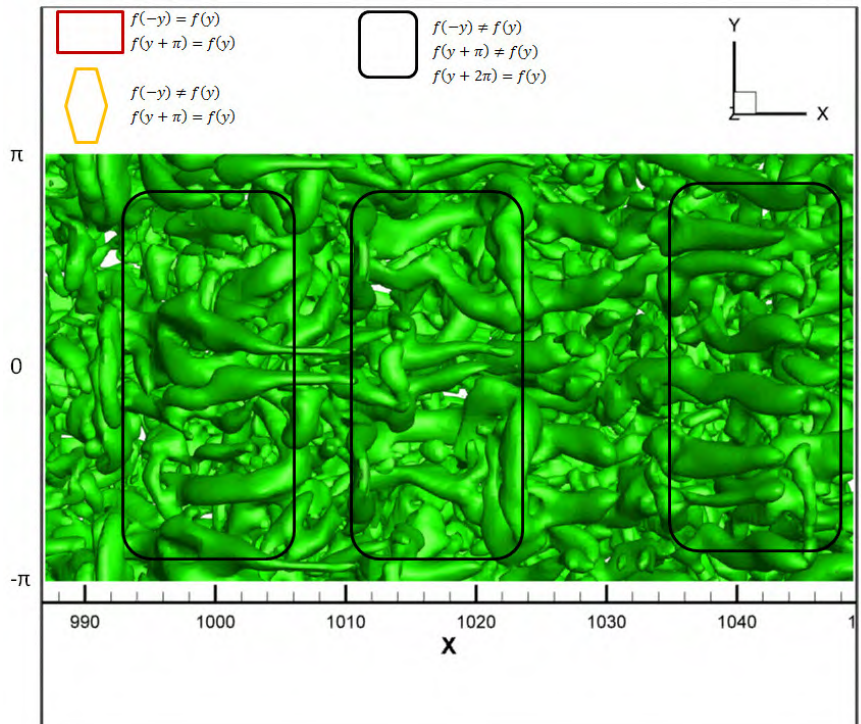

Figure 22(c): Periodic but asymmetric at t=28.1T – all black boxes: periodic but asymmetric (period= 2π); the spanwise domain is –π<y<π

16
American Institute of Aeronautics and Astronautics

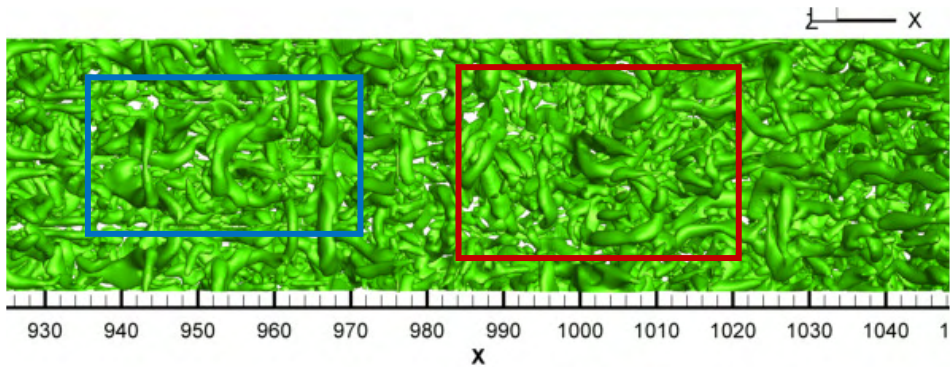

(a) Top ring structure lost symmetry (blue area is symmetric but red area is not)

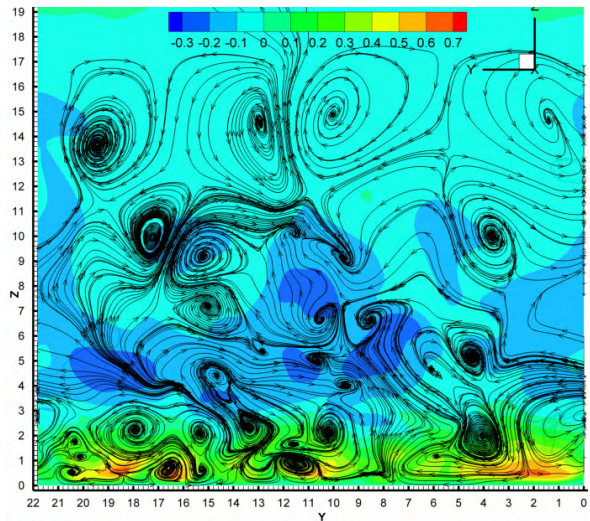

(b) Symmetry loss in the whole section of y-z plane

Figure 23: The whole flow field lost symmetry at t=21.25T

## V. Hypothesis

Apparently, loss of symmetry, which is equivalent to adding sin (2ky), is a key issue for the starting point of flow randomization. A hypothesis was given by Liu that the loss of symmetry is caused by the shift from C-type to K-type transitions or reverse. The randomization is caused by the instability of the multiple level vortex ring cycle structure which is an internal property. Therefore, randomization is not mainly caused by big background noises or removal of periodic boundary conditions in DNS. The flow shows a C-typed transition (staggered) at beginning (Figure 24(a)), but becomes K-type transition later (Figure 24(b)) and then mixed (Figure 24(c)). This means the first vortex circle is C-type, but second circle, which overlaps the first circle, is K-type. K-type, C-type, and mixed type are all observed by experiment. There must be some trend of shift from C-type to K-type or reverse. This shift will cause the loss of symmetry of the middle of the first circle (Figure 25). This trend will change the underneath large ring structure and cause the loss of symmetry of the underneath large rings. Once the middle large rings lost symmetry, the underneath small length scale will lose the symmetry quickly due to the asymmetry of second sweeps. Finally, the asymmetry of lower level vortex structures will affect the top rings through ejection and the whole flow field becomes turbulent.

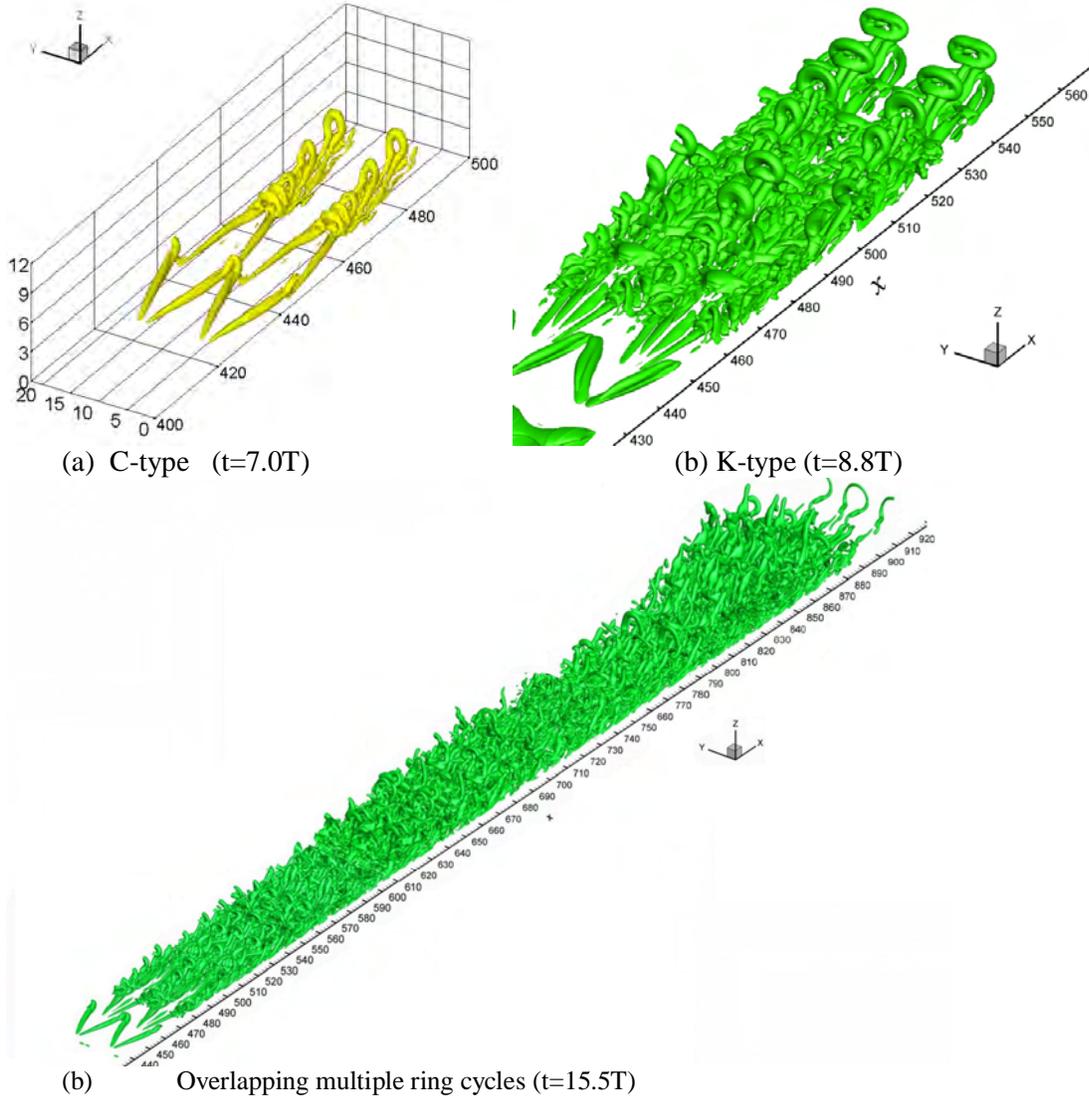

(a) C-type (t=7.0T)  (b) K-type (t=8.8T)

(b) Overlapping multiple ring cycles (t=15.5T)
Figure 24: Vortex structure in K-type, C-type and mixed type transition (T is the T-S period)

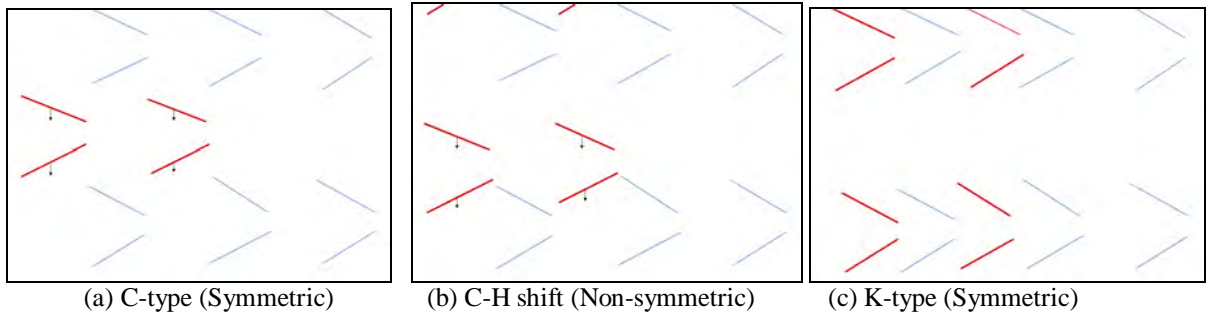

(a) C-type (Symmetric)　　(b) C-H shift (Non-symmetric)　　(c) K-type (Symmetric)

Figure 25: Sketch of symmetry loss due to the shift from C-type transition (vortex circles are staggered) to the K-type transition (vortex circles are aligned) (the blue one is the first circle and the red one is the second circle)

## VI. Conclusion

Although we still use the symmetric boundary condition (period is π for inflow and 2π for the whole domain) without intentional introduction of background noises from the inflow, outflow, and far-field, we still find that the vortices become asymmetric in the whole flow field finally. Following conclusions can be made by the current DNS result we obtained:

1. The phenomenon of asymmetry is first observed at the middle level of the overlapping multiple ring cycles instead of the ring tip.
2. The loss of flow symmetry is also found at the middle part of the flow field in the streamwise direction. Meanwhile the downstream and inflow parts still keep symmetric characteristics.
3. Since there are small vertex rings generated at the middle by different streamwise velocity shear levels, they will affect the intensity of positive spikes. This will result in deformation of the small vortices near the bottom of the boundary layer.
4. The asymmetric lower level vortices deform the shape of the upper level vortices through ejection.
5. Finally, we can find that the top flow structure loses the symmetry and the whole flow field is randomized.

In summary, the order of asymmetric phenomenon is: middle ring structure first, bottom later and top last.

## Acknowledgments

This work was supported by AFOSR grant FA9550-08-1-0201 supervised by Dr. John Schmisseur and the Department of Mathematics at University of Texas at Arlington. The authors are grateful to Texas Advanced Computing Center (TACC) for providing computation hours. This work is accomplished by using Code DNSUTA which was released by Dr. Chaoqun Liu at University of Texas at Arlington in 2009.

[11] Guo, Ha; Borodulin, V.I..; Kachanov, Y.s.; Pan, C; Wang, J.J.; Lian, X.Q.; Wang, S.F., Nature of sweep and ejection events in transitional and turbulent boundary layers, J of Turbulence, August, 2010

[12] Herbert, T., 1988, "Secondary Instability of Boundary Layer," Annu. Rev.Fluid Mech., 20, pp. 487-526.

[13] Jeong J., Hussain F. On the identification of a vortex, J. Fluid Mech. 1995, 285:69-94

[14] Jiang, L., Chang, C. L. (NASA), Choudhari, M. (NASA), Liu, C., Cross-Validation of DNS and PSE Results for Instability-Wave Propagation, AIAA Paper #2003-3555, The 16th AIAA Computational Fluid Dynamics Conference, Orlando, Florida, June 23-26, 2003

[15] Kachnaov, Y. S., 1994, "Physical Mechanisms of Laminar-Boundary-Layer Transition," Annu. Rev. Fluid Mech., 26, pp. 411–482.

[16] Kachanov, Y.S. On a universal mechanism of turbulence production in wall shear flows. In: Notes on Numerical Fluid Mechanics and Multidisciplinary Design. Vol. 86. Recent Results in Laminar-Turbulent Transition. — Berlin: Springer, 2003, pp. 1–12.

[17] Kleiser L, Zang T A. Numerical simulation of transition in wall-bounded shear flows. Annu.Rev.Fluid Mech.1991.23:495-537

[18] Lee C B., Li R Q. A dominant structure in turbulent production of boundary layer transition. Journal of Turbulence, 2007, Volume 8, N 55

[19] Liu, X., Chen, L., Oliveira, M., Tang, D., Liu, C., DNS for late stage structure of flow transition on a flat-plate boundary layer, AIAA Paper 2010-1470, Orlando, FL, January 2010a.

[20] Liu, C., Chen, L., Study of mechanism of ring-like vortex formation in late flow transition, AIAA Paper 2010-1456, Orlando, FL, January 2010b.

[21] Liu, X., Chen, Z., Liu, C., Late-Stage Vortical Structures and Eddy Motions in Transitional Boundary Layer Status, Chinese Physics Letters Vol. 27, No.2 2010c

[22] Liu, C., Chen, L., Lu, P., New Findings by High Order DNS for Late Flow Transition in a Boundary Layer, J of Modeling and Simulation in Engineering, to appear, open access journal and copy right is kept by authors, 2011a

[23] Liu, C., Chen, L., Parallel DNS for vortex structure of late stages of flow transition, J. of Computers and Fluids, Vol.45, pp 129–137, 2011b

[24] Liu, C., Numerical and Theoretical Study on "Vortex Breakdown", International Journal of Computer Mathematics, to appear, 2011c

[25] Liu, C., Chen, L., Lu, P., and Liu, X., Study on Multiple Ring-Like Vortex Formation and Small Vortex Generation in Late Flow Transition on a Flat Plate, Theoretical and Numerical Fluid Dynamics, to appear, 2011d

[26] Lu, P., Liu, C., Numerical Study of Mechanism of U-Shaped Vortex Formation, AIAA Paper 2011-0286, and Journal of Computers and Fluids, to appear, 2011a

[27] Lu, P. and Liu, C., DNS Study on Mechanism of Small Length Scale Generation in Late Boundary Layer Transition, AIAA Paper 2011-0287 and J. of Physica D, Non-linear, to appear, 2011b, on line: http://www.sciencedirect.com/science/article/pii/S0167278911002612

[28] Marshak, Alex, *3D radiative transfer in cloudy atmospheres; pg.76*. Springer. ISBN 9783540239581. http://books.google.com/books?id=wzg6wnpHyCUC, 2005

[29] Meyer, D.G.W.; Rist, U.; Kloker, M.J. (2003): Investigation of the flow randomization process in a transitional boundary layer. In: Krause, E.; Jäger, W. (eds.): *High Performance Computing in Science and Engineering '03.* Transactions of the HLRS 2003, pp. 239-253 (partially coloured), Springer.

[30] Moin, P., Leonard, A. and Kim, J., Evolution of curved vortex filament into a vortex ring. *Phys. Fluids*, **29**(4), 955-963, 1986

[31] Mullin, Tom, Turbulent times for fluids, *New Scientist*., 11 November 1989

[32] Rist, U., et al. Turbulence mechanism in Klebanoff transition: a quantitative comparison of experiment and direct numerical simulation. J. Fluid Mech. 2002, 459, pp. 217-243.

[33] Rist, U. and Kachanov, Y.S., 1995, Numerical and experimental investigation of the K-regime of boundary-layer transition. In: R. Kobayashi (Ed.) *Laminar-Turbulent Transition* (Berlin: Springer) pp. 405-412.
20
American Institute of Aeronautics and Astronautics